\documentclass[preprint,showpacs,preprintnumbers,amsmath,amssymb]{revtex4}
\newtheorem{theorem}{\indent Theorem}
\newtheorem{lemma}[theorem]{\indent Lemma}

\newtheorem{algorithm}{\indent Algorithm}
\newtheorem{remark}{Remark}
\newenvironment{proof}{\par{\itshape Proof}.\ }
    {\hfill\raisebox{.56ex}{\fbox{}}\par}

\usepackage{graphicx}
\usepackage{dcolumn}
\usepackage{bm}

\begin{document}


\title{Incoherent Control of Locally Controllable Quantum Systems}

\author{Daoyi  Dong$^{1,2}$}
 \email{dydong@amss.ac.cn}
\affiliation{%
1. Institute of Cyber-Systems and Control, National Laboratory of
Industrial Control Technology, Zhejiang University, Hangzhou 310027, China \\
2. Institute of Systems Science, AMSS, Chinese Academy of
Sciences, Beijing 100190, China\\
}%

\author{Chenbin Zhang}
\affiliation{ Department of Automation, University of Science
and Technology of China, Hefei 230027, China\\
}%

\author{Herschel Rabitz}
\author{Alexander Pechen}%
\affiliation{%
Department of Chemistry, Princeton University,
Princeton, New Jersey 08544 USA\\
}%

\author{Tzyh-Jong Tarn}
\affiliation{ Department of Electrical and Systems Engineering,
Washington University in St. Louis, St. Louis, MO 63130 USA\\
}%

\date{\today}

\begin{abstract}
An incoherent control scheme for state control of locally
controllable quantum systems is proposed. This scheme includes three
steps: (1) amplitude amplification of the initial state by a
suitable unitary transformation, (2) projective measurement on the
amplified state, and (3) final optimization by a unitary controlled
transformation. The first step increases the amplitudes of some
desired eigenstates and the corresponding probability of observing
these eigenstates, the second step projects, with high probability,
the amplified state into a desired eigenstate, and the last step
steers this eigenstate into the target state. Within this scheme,
two control algorithms are presented for two classes of
quantum systems. As an example, the incoherent
control scheme is applied to the control of a hydrogen atom by an
external field. The results support the suggestion that projective
measurements can serve as an effective control and local
controllability information can be used to design control laws for
quantum systems. Thus, this scheme establishes a subtle connection
between control design and controllability analysis of quantum
systems and provides an effective engineering approach for
controlling quantum systems with partial controllability
information.
\end{abstract}

\pacs{42.50.Dv, 02.30.Yy, 03.65.-w}
\maketitle

\section{Introduction}
Control of quantum phenomena plays an important role in various research
fields including physical chemistry~\cite{Rabitz et al 2000,Chen et al
2005,Rabitz 2003,Shapiro and Brumer 2003,Dantus 2004}, atomic and molecular
physics~\cite{Judson and Rabitz 1992,Chu 2002,Bonacic 2005}, quantum
information~\cite{Nielsen and Chuang 2000,Grace et al 2007:a,Grace et al
2007:b} and future quantum technologies~\cite{Dowling and Milburn 2003}. In
quantum control theory, the controllability of quantum systems is
the first fundamental issue to address~\cite{Huang et al 1983} due to its
practical importance, including a close connection with the universality of
quantum computation~\cite{Ramakrishna and Rabitz 1996}, the possibility of
attaining atomic or molecular scale transformations~\cite{Ramakrishna et al 1995,Wu et al 2006}, etc. Different notions of controllability exist including pure state controllability,
complete controllability, wavefunction controllability, and kinematic controllability in the set of all density matrices~\cite{Ramakrishna et
al 1995}-\cite{Wu et al 2007}. A common research focus is on finite
dimensional quantum systems for which the controllability criteria may be expressed
in terms of the structure and rank of the corresponding Lie groups and Lie
algebras~\cite{Chen et al 2005}. This method allows for the easy mathematical
treatment of closed quantum systems and in some cases can directly benefit from classical
control theory. Various results have been derived for specific problems using this
method~\cite{Albertini and D'Alessandro 2003,Schirmer et al 2001}, but the relevant criteria may be computationally difficult when the
dimension of the controlled system is large. Turinici and
Rabitz~\cite{Turinici and Rabitz 2001,Turinici and Rabitz 2003} proposed a
wavefunction controllability method based on graph theory, and its
controllability criterion becomes easy to verify.

Most of the existing results consider control within the whole state
space of controlled quantum systems, and we call this global
controllability. In practical applications, some quantum systems may
not be globally controllable, or the information about global
controllability may be difficult to acquire. However, it may be easy
to obtain local controllability information. For example,
Beauchard~\cite{Beauchard 2005} proved that a nonrelativistic
charged particle in a 1-D box controlled by a uniform electric field
is locally controllable around the ground state. Furthermore,
Beauchard and Coron~\cite{Beauchard and Coron 2006} also proved that
in some cases two eigenstate locally controllable wavefunctions can
be moved exactly from one to another in a finite time. Local
controllability has also been an interesting topic for classical
mechanical systems~\cite{Khapalov 2007} and in chemical reaction
control~\cite{Farkas 1998}. Moreover, in some situations, the focus
may only be on the control within some state subspace of the
controlled system. For example, in quantum computation one may only
focus on decoherence-free subspaces (DFS)~\cite{Lidar et al
1998,Kwiat et al 2000,Kiffner et al 2007}. The control design of
locally controllable quantum systems is very relevant for such
problems and in this paper we focus on quantum systems with local
controllability information.

The effective determination of control strategies is another important problem in
quantum control theory. The main paradigm is coherent control where one
manipulates the state of the system by applying a semiclassical
potential in a fashion that preserves quantum coherence \cite{Lloyd
2000}-\cite{Rice and Zhao 2000}. This approach has successfully been used, e.g., for
control of chemical reactions via two-beam interference control and two-pulse
time delay control~\cite{Rice 2001,Rice and Zhao 2000}. The method of learning control has been developed~\cite{Judson and Rabitz 1992}, which has the essential advantage of being applicable to control problems lacking detailed information about the controlled system and its dynamics.

In coherent control, quantum measurements are commonly viewed as having
a deleterious effect, since they destroy the coherent state of the measured system.
However, recent results show that in some situations quantum measurements can
be beneficial for quantum control, and they can be combined with unitary operators to
achieve certain quantum control tasks or even can make nonunitarily
controllable systems controllable~\cite{Vilela Mendes and Man'ko
2003}-\cite{Shuang et al 2007}. In these schemes quantum measurements destroy
the coherent characteristics of the controlled systems, and thus such control can be called ``incoherent control". Various methods for direct incoherent control by
the environment were also proposed~\cite{Pechen and Rabitz 2006,Pechen and
Rabitz 2008,Romano and D'Alessandro 2006-1,Romano and D'Alessandro 2006-2}.

In this paper, we propose an incoherent control scheme for locally
controllable quantum systems. This scheme can be considered as a
specific type of measurement-assisted control~\cite{Pechen et al
2006,Shuang et al 2007} and includes three steps: (1) amplitude
amplification of the initial state, (2) projective measurement on
the amplified state, and (3) subsequent transfer of the resultant
state into the target state. The first and the last steps are
realized by unitary operators, and thus correspond to coherent
control. The intermediate step is realized by a measurement, and
thus corresponds to incoherent control. The first step enhances the
probability of success by enhancing the amplitude of some desired
system eigenstates. The second step projects, with high probability
(the probability of success), the amplified state onto a desired
eigenstate. The last step transfers the desired eigenstate into the
target state. This quantum control method is probabilistic due to
the nature of quantum measurements; however, the probability of
success can be greatly enhanced by amplitude amplification
technology~\cite{Brassard et al 1998,Hoyer 2000,Grover 1998}. Within
this scheme, we present two incoherent control algorithms: one for a
target state which is reachable from a specific
eigenstate, and the other for quantum systems with wavefunction
controllable subspaces. As an example, we investigate the control of
a hydrogen atom subject to an external field. The results
demonstrate that projective measurement and local controllability
information can be very helpful for the control design of quantum
systems.

The paper is organized as follows. Section II introduces the notions
of local controllability and a wavefunction controllable subspace.
In Section III, the two algorithms based on the incoherent control
scheme with quantum amplitude amplification and projective
measurement are formulated for two classes of quantum systems.
An illustration is given to demonstrate the incoherent control scheme in Section IV. Concluding remarks are given in Section V.

\section{Local controllability and a wavefunction controllable subspace}
This section first describes the control model for general finite-level
quantum systems. Then we give the definitions of local and global
controllability using the notion of a reachable set. Finally, we introduce the
concepts of a wavefunction controllable subspace and wavefunction
controllability, considering quantum systems with wavefunction controllable
subspaces as a special class of locally controllable systems.

\subsection{Quantum Control Model}
In quantum mechanics, the state  of a closed quantum system at time $t$ can be
represented by a vector $|\psi(t)\rangle$ in some
Hilbert space $\mathcal{H}$. The state $|\psi(t)\rangle$ of a closed quantum
system evolves according to the Schr\"{o}dinger equation \cite{Turinici and
Rabitz 2003}
\begin{equation}\label{schrodinger}
\iota\hbar\frac{\partial}{\partial
t}|\psi(t)\rangle=H_{0}|\psi(t)\rangle,\qquad
|\psi(t=0)\rangle=|\psi_{0}\rangle
\end{equation}
where $\iota=\sqrt{-1}$, $H_{0}$ is the internal Hamiltonian of the
system (i.e., a Hermitian operator in $\cal H$), $\hbar$ is Planck's
constant, and the initial state has unit norm $\|\psi_0\|^2\equiv
\langle\psi_{0}|\psi_{0}\rangle=1$. The control of the system is
realized by a control function $u(t)\in L^2(\mathbf{R})$ coupled to
the system via a time-independent Hermitian interaction Hamiltonian
$H_{I}$ (e.g., dipole moment coupling). The total Hamiltonian
$H=H_{0}+u(t)H_{I}$ determines the controlled evolution
\begin{equation}\label{controlled model}
\iota\hbar\frac{\partial}{\partial t}|\psi_{u}(t)\rangle
=[H_{0}+u(t)H_{I}]|\psi_{u}(t)\rangle.
\end{equation}
The goal of the control is to find a final time $T>0$ and a finite
energy input $u(t)\in L^{2}([0, T],\mathbf{R})$ which drives the
system from the initial state $|\psi_{0}\rangle$ into some
predefined target state $|\psi_{\text{target}}\rangle$ (e.g.,
population transfer using external control fields).

An important problem to assess is the controllability of the system
which establishes which states can be connected by the admissible
controls. The controllability analysis for infinite dimensional
quantum systems is typically difficult to assess. For
simplification, we consider finite dimensional quantum systems,
which is an appropriate approximation in many practical situations.
For a finite dimensional quantum system, $\cal H$ is a finite
dimensional Hilbert space, the free Hamiltonian $H_0$ is a Hermitian
operator in $\cal H$ and its set of eigenstates
$D=\{|\phi_{i}\rangle;i=1,\dots,N\}$ forms a basis in $\cal H$. The
evolving state $|\psi_{u}(t)\rangle$ can be expanded in this basis
as
\begin{equation}\label{superposition}
|\psi_{u}(t)\rangle=\sum_{i=1}^{N}c_{i}(t)|\phi_{i}\rangle.
\end{equation}
The total Hamiltonian $H=H_{0}+u(t)H_{I}$ defines a unitary
evolution operator (propagator) $U(u,t_1\to t_2)$ such that for any
state $|\psi_{1}\rangle$ the state $U(u,t_{1}\rightarrow
t_{2})|\psi_{1}\rangle$ is the solution at time $t=t_{2}$ of
equation (\ref{controlled model}) with the initial state
$|\psi_{1}\rangle$ at time $t=t_{1}$. In particular,
\begin{equation}\label{evolution}
|\psi_{u}(t)\rangle=U(u,0\to t)|\psi_{0}\rangle,
\end{equation}
Unitarity of the propagator $U(u,0\to t)$ implies that the state
$|\psi_{u}(t)\rangle$ evolves on the complex unit sphere
$S_{\mathbf{C}}^{N-1}:=\{|\psi\rangle\in{\cal H}\ :\ \|\psi\|=1\}$.
Substitution of Eq. (\ref{superposition}) into Eq. (\ref{controlled
model}) gives the following equation for the coefficients
$C(t)=\{c_i(t)\}_{i=1}^N$~\cite{Turinici and Rabitz 2003}:
\begin{eqnarray}
\iota\hbar\frac{\partial C(t)}{\partial t}&=&[A+u(t)B]C(t),\quad C(t=0)=C_{0},\label{model}\\
C_{0}&=&(c_{0i})_{i=1}^{N},\ \
c_{0i}=\langle\phi_{i}|\psi_{0}\rangle,\ \
\sum_{i=1}^{N}|c_{0i}|^{2}=1\quad \label{coefficient}
\end{eqnarray}
Here $A$ and $B$ are the matrices of the operators $H_{0}$ and
$H_{I}$, respectively, in the basis $|\phi_i\rangle$. The $A$ matrix
is diagonal and the $B$ matrix is Hermitian~\cite{Turinici and
Rabitz 2001}, and in order to avoid the trivial control problem, we
assume $[A, B]:=AB-BA\neq 0$.

\subsection{Local and global controllability}
In the development of quantum control theory, the controllability of
quantum systems has been analyzed from different perspectives. Huang
\emph{et al.}~\cite{Huang et al 1983} studied the controllability of
quantum systems using Nelson's analytic domain theory. Ramakrishna
\emph{et al.} \cite{Ramakrishna et al 1995} investigated
controllability and proposed a simple algorithm
to detect their controllability. Schirmer \emph{et al.}
\cite{Schirmer et al 2001} obtained a sufficient condition for
complete controllability of $N$-level quantum systems subject to a
particular control pulse. Turinici and Rabitz \cite{Turinici and
Rabitz 2001,Turinici and Rabitz 2003} investigated the
controllability of quantum systems using graph theory and proved
exact wavefunction controllability of finite dimensional models
under very natural hypotheses. Albertini and D'Alessandro
\cite{Albertini and D'Alessandro 2003} defined several different
notions of controllability for multilevel quantum systems and
established some connections among these different notions. Altafini
\cite{Altafini 2003} studied the controllability properties for
finite dimensional quantum Markovian master equations. Wu \emph{et
al.} \cite{Wu et al 2006} considered the smooth controllability of
infinite-dimensional quantum-mechanical systems. Wu \emph{et
al.}~\cite{Wu et al 2007} studied controllability of open quantum
system subject to arbitrary Kraus-type dynamics and showed that such
systems are completely density matrix controllable.

Most existing results consider controllability in the whole
state space of the quantum system, i.e., global controllability, such as complete
controllability \cite{Schirmer et al 2001}, pure state
controllability \cite{Albertini and D'Alessandro 2003}, and
wavefunction controllability \cite{Turinici and Rabitz 2003}.
This paper considers the control problem for a class of
quantum systems with local controllability information. First, we
will define local and global controllability using the concept
of reachable sets. We say that a state $|\psi_{2}\rangle$ is reachable
from $|\psi_{1}\rangle$ at time $0<T<\infty$ if there exists a
finite energy control $u(t)\in L^{2}([0,T];\mathbf{R})$ such that
$U(u,0\rightarrow T)|\psi_{1}\rangle=|\psi_{2}\rangle$. The
reachable set $R(|\psi\rangle)$ of a state $|\psi\rangle$ is the set
of all states reachable from $|\psi\rangle$. If the reachable set
covers the whole state space $S_{\mathbf{C}}^{N-1}$, the
corresponding circumstance entails global controllability, that
is to say, a quantum system is globally controllable if all its
states are controllable in the whole state space
$S_{\mathbf{C}}^{N-1}$ and therefore
$R(|\psi\rangle)=S_{\mathbf{C}}^{N-1}$ for any $|\psi\rangle \in
S_{\mathbf{C}}^{N-1}$. If $|\psi_{i}\rangle$ is an eigenstate
$|\phi_{g}\rangle$ of $H_0$ and $|\psi_{\rm target}\rangle \in
R(|\psi_{i}\rangle)$, we say that $|\psi_{\rm target}\rangle$
is eigenstate reachable. This situation with eigenstate reachable target states corresponds to local controllability and will be considered in this paper. In
order to avoid a triviality, we assume
$|\psi_{\text{target}}\rangle\neq |\phi_{g}\rangle$. As discussed in
the Introduction, the assumption that some target states are
eigenstate reachable is reasonable for various practical
applications. Before presenting our incoherent control strategy for
locally controllable systems, we will introduce a special class of
locally controllable systems which have a wavefunction controllable
subspace.

\subsection{Wavefunction controllability and the wavefunction controllable subspace}\label{WC and WCS}
To study the controllability of the system with the evolution
equation~(\ref{model}), Turinici and Rabitz \cite{Turinici and
Rabitz 2001,Turinici and Rabitz 2003} associate to the system a
non-oriented connectivity graph $G=(V,E)$, where the set $V$ of
vertices consists of the eigenstates $|\phi_{i}\rangle$ and the set
of edges $E$ consists of all pairs of eigenstates directly coupled
by the matrix $B$,
\begin{eqnarray}\label{graph}
G(V,E):&&V=\{|\phi_{1}\rangle,\dots,|\phi_{N}\rangle\},\nonumber\\
&&E=\{(|\phi_{i}\rangle,|\phi_{j}\rangle);i<j, B_{ij}\neq 0\} \ .
\end{eqnarray}
Let $G_{k}=(V^{(k)},E^{(k)}), k=1,\dots,K$ be connected components
of this graph. Denote by $\lambda_{i}$ $(i=1,\dots, N)$ the
eigenvalues of the matrix $A$ and let
$\nu_{ij}=\lambda_{i}-\lambda_{j}$ ($i,j=1,\dots,N$). The following
lemma provides the criteria for wavefunction controllability in
terms of the connectivity graph~\cite{Turinici and Rabitz
2001,Turinici and Rabitz 2003}:

\begin{lemma}[wavefunction controllability]\label{WC}
The system (\ref{model}) is wavefunction controllable if the following
assumptions hold:\\
(I) The graph $G$ is connected, i.e. $K=1$.\\
(II) The graph $G$ does not have ``degenerate transitions", that is
for all $(i,j)\neq (a,b)$, $i\neq j$, $a\neq b$ such that
$B_{ij}\neq 0$, $B_{ab}\neq
0$: $\nu_{ij}\neq \nu_{ab}$.\\
(III) For each $i,j,a,b=1,\dots,N$ such that $\nu_{ij}\neq 0$ the
number $(\nu_{ab}/\nu_{ij})$ is rational.
\end{lemma}
The proof of Lemma \ref{WC} can be found in~\cite{Turinici and Rabitz 2003}.
The three assumptions in Lemma \ref{WC} provide a sufficient but not a
necessary condition for wavefunction controllability. In some circumstances,
the assumptions (II) and (III) can be slightly relaxed~\cite{Turinici and
Rabitz 2001}.

For some practical systems, the assumption~(I) of Lemma~\ref{WC} may
not be satisfied or verifying this assumption may be difficult. In
such cases considering the problem of control inside an appropriate
state subspace of the quantum system may be more practical.
Moreover, for some controlled quantum systems partial
controllability information can be relatively easy to obtain from
some physical and chemical experiments. Assuming $K\neq1$, let
$G_{\omega}$ be a fixed connected component of $G(V,E)$ with the set
of vertices denoted as $D_{\omega}=\{|\phi_{1}^{'}\rangle,\dots,
|\phi_{M}^{'}\rangle\}$ ($1<M< N$). Obviously, $G_{\omega}\subset
G(V,E)$ and $D_{\omega}\subset D$. Denoting the subspace generated
by $D_{\omega}$ as $\Omega$, we have the following
theorem~\cite{Dong et al 2007}:
\begin{theorem}[wavefunction controllable subspace]\label{WCS}
The subspace $\Omega$ is wavefunction controllable if
the
following assumptions hold: \\
(I) The graph $G_{\omega}$ does not have ``degenerate transitions",
that is for all $(k,l)\neq (\alpha,\beta)$, $k\neq l$, $\alpha\neq
\beta$ such that $B_{kl}\neq 0$,
$B_{\alpha\beta}\neq 0$: $\nu_{kl}\neq \nu_{\alpha\beta}$. \\
(II) For each $k,l,\alpha,\beta=1,\dots,M$ such that $\nu_{kl}\neq
0$ the number $(\nu_{\alpha\beta}/\nu_{kl})$ is rational.
\end{theorem}
\noindent\begin{proof} Since $G_{\omega}$ is a connected component
of $G(V,E)$, it is a connected graph itself. The subspace $\Omega$
is generated by the set $D_{\omega}$ of all vertices of $G_{\omega}$
and therefore Lemma~\ref{WC}, under the assumptions (I) and (II),
immediately implies that $\Omega$ is a wavefunction controllable
subspace.
\end{proof}

As mentioned above, the property of wavefunction controllability of the system
can be viewed as a type of global controllability, whereas the existence of a
wavefunction controllable subspace can be considered as a form of local
controllability. That is, the states in the wavefunction controllable subspace
are locally controllable. In the following sections, besides the case when the
target state is reachable from a specific eigenstate, we will also
investigate the case when the target state belongs to some wavefunction
controllable subspace.

\section{Incoherent control of locally controllable quantum systems}
This section describes the proposed incoherent quantum control
scheme. First, we briefly discuss general coherent and incoherent
quantum control methods. Then we present quantum amplitude
amplification technology which plays an important role in the
proposed incoherent control scheme. Finally we formulate two
algorithms for the incoherent control scheme: one is designed for
general eigenstate reachable target states and the other for
quantum systems with a wavefunction controllable subspace as
discussed in Sec.~\ref{WC and WCS}.

\subsection{Incoherent control}
The determination of suitable control strategies is an important objective in
quantum control theory. Coherent control is a particularly powerful quantum
control strategy, where the system is controlled by applying a
semiclassical potential~\cite{Lloyd 2000}. The controls commonly appear as the
tunable parameters in the Hamiltonian of the system, which can directly affect the coherent
part of the system's dynamics. The early
paradigms of quantum control mainly concentrated on the open loop coherent
control strategy~\cite{Judson and Rabitz 1992,Warren et al 1993} which has achieved wide success, e.g., in control of chemical reactions~\cite{Rice 2001}. Learning based coherent control~\cite{Judson and Rabitz 1992} is proving to be extremely useful for manipulating quantum systems even when detailed information about their structure, coupling to the control field, etc. is not available. In recent years coherent control has also been proposed for incorporation into quantum feedback control, where
measurements on the system are used to determine its state and then the measured outcome is exploited to generate a semiclassical potential applied to the system to coherently guide it into a desired state~\cite{Lloyd 2000}.

Quantum measurements generally destroy the coherent characteristics
of the quantum system, and without management that circumstance can
create difficulties for attaining successful control. However, in
some circumstances quantum measurements can be effectively used for
controlling quantum systems~\cite{Vilela Mendes and Man'ko
2003,Mandilara and Clark 2005,Gong and Rice 2004,Sugawara 2005,Roa
et al 2006, Pechen et al 2006, Shuang et al 2007}. For example,
Vilela Mendes and Man'ko~\cite{Vilela Mendes and Man'ko 2003} showed
that under suitable conditions a system that is not controllable
under unitary transformations can become controllable under the
joint action of projective measurements and unitary evolution. Gong
and Rice~\cite{Gong and Rice 2004} found that measurements can
assist a coherent strategy to control the population transfer
branching ratio between degenerate product states.
Sugawara~\cite{Sugawara 2005} studied quantum dynamics driven by
continuous laser fields under the measurement process and explored
the possibility of measurement-assisted quantum control. Mandilara
and Clark~\cite{Mandilara and Clark 2005} proposed a probabilistic
quantum control scheme via indirect measurement. Roa \emph{et al.}
\cite{Roa et al 2006} applied sequential measurements of two
noncommuting observables to drive an unknown mixed quantum state to
a known pure state without the use of unitary transformations.
Rabitz and co-workers explored the use of nonselective von Neumann
measurements in conjunction with optimal control to enhance the
capability of controlling quantum systems~\cite{Pechen et al
2006,Shuang et al 2007}. Incoherent control by manipulating the
environment was also considered~\cite{Pechen and Rabitz 2006,Pechen
and Rabitz 2008,Romano and D'Alessandro 2006-1,Romano and
D'Alessandro 2006-2}. Pechen and Rabitz~\cite{Pechen and Rabitz
2006,Pechen and Rabitz 2008} presented a general method of
incoherent control by the environment, where tailored incoherent
radiation or a medium (e.g., atomic or molecular gas, solvent, etc.)
is used as a control tool to drive the system to a target state via
non-unitary evolution. Romano and D'Alessandro~\cite{Romano and
D'Alessandro 2006-1,Romano and D'Alessandro 2006-2} considered an
incoherent control scheme where optimization of the state of an
ancillary finite-level system is used to drive the controlled system
to an arbitrary target state. We refer to all of these schemes as
``incoherent control'', since the coherent characteristics of the
controlled system are destroyed in the process. Along this line, we
propose an incoherent control scheme for a class of quantum systems
where the target state is eigenstate reachable. This
scheme includes three basic steps. In the first step quantum
amplitude amplification of the initial state is used to increase the
probability of success. In the second step a measurement is used to
project, with high probability, the amplified state into a desired
eigenstate. Finally, a controlled unitary transformation is used to
steer this eigenstate into the target state. We will first introduce
the quantum amplitude amplification concept and then describe the
incoherent control algorithms.

\subsection{Quantum amplitude amplification}\label{QAA}
Quantum amplitude amplification is a powerful ingredient in quantum
algorithms~\cite{Brassard et al 1998,Hoyer 2000}. It is a natural
generalization of Grover's quantum search algorithm, which allows
for a speedup of many classical algorithms~\cite{Brassard et al
1998}-\cite{Grover 1998}. Amplitude amplification was first used by
Brassard and H{\o}yer \cite{Brassard and Hoyer 1997} to construct an
exact quantum polynomial-time algorithm for solving Simon's problem.
The central task of quantum amplitude amplification is to find a
suitable operator $\mathbf{Q}$ whose iterative action on the initial
state can increase the probability of success roughly by a constant
at each iteration, in analogy to the iteration process in
probabilistic algorithms~\cite{Brassard et al 1998}.

Let $|\Phi\rangle$ be a pure state of an $N$-level quantum system.
This state can be represented as a superposition of some orthonormal
basis states $\mathbb{X}=\{|0\rangle, \dots, |x\rangle, \dots,
|N-1\rangle\}$ in the $N$-dimensional system Hilbert space
$\mathcal{H}$, i.e., $|\Phi\rangle=\sum_{x=0}^{N-1}c_{x}|x\rangle$,
where $\sum_{x=0}^{N-1}|c_{x}|^{2}=1$. A Boolean function
$\chi:\mathbb{X}\rightarrow\{0,1\}$ induces two orthogonal subspaces
of $\mathcal{H}$: the ``good" subspace and the ``bad" subspace. The
good subspace is spanned by the set of basis states $|x\rangle\in
\mathbb{X}$ satisfying $\chi(x)=1$ and the bad subspace is its
orthogonal complement in $\mathcal{H}$. We denote by $P_g$ the
projector onto the ``good" subspace. Every pure state $|\Phi\rangle$
in $\mathcal{H}$ can be decomposed as
$|\Phi\rangle=|\Phi_g\rangle+|\Phi_b\rangle$, where
$|\Phi_g\rangle=P_g|\Phi\rangle$ denotes the projection of
$|\Phi\rangle$ onto the good subspace and $|\Phi_b\rangle=(\mathbb
I-P_g)|\Phi\rangle$ denotes the projection of $|\Phi\rangle$ onto
the bad subspace (here $\mathbb I$ is the identity operator).
According to quantum measurement theory, the occurrence
probabilities of the ``good" state $|x\rangle$ [$\chi(x)=1$] and the
``bad" state $|x\rangle$ [$\chi(x)=0$] upon measuring $|\Phi\rangle$
are $g=\langle\Phi_g|\Phi_g\rangle$ and
$b=\langle\Phi_b|\Phi_b\rangle=1-g$, respectively.

Let $\mathcal{U}$ be a quantum algorithm that acts in $\mathcal{H}$
without measurements (i.e., $\mathcal{U}$ is a unitary operator) and
let $|\Phi\rangle=\mathcal{U}|0\rangle$. Given two angles $0\leq
\varphi_{1}, \varphi_{2} \leq \pi$, a general quantum amplitude
amplification can be realized by the following operator
\cite{Brassard et al 1998}
\begin{equation}
\mathbf{Q}=\mathbf{Q}(\mathcal{U},\chi,\varphi_{1},\varphi_{2})
=-\mathcal{U}\mathcal{P}_{0}^{\varphi_{1}}\mathcal{U}^{-1}\mathcal{P}_{\chi}^{\varphi_{2}}
\end{equation}
The operators $\mathcal{P}_{0}^{\varphi_{1}}$ and
$\mathcal{P}_{\chi}^{\varphi_{2}}$ conditionally change the phase of
the amplitudes of state $|0\rangle$ and the good states respectively
\cite{Brassard et al 1998}:
\begin{eqnarray}
\mathcal{P}_{0}^{\varphi_{1}}|x\rangle &=& \left\{%
\begin{array}{ll}
    e^{\iota\varphi_{1}}|x\rangle, & \hbox{if  $x=0$;} \\
    |x\rangle, & \hbox{if  $x\neq0$.} \\
\end{array}%
\right.\\
\mathcal{P}_{\chi}^{\varphi_{2}}|x\rangle &=& \left\{%
\begin{array}{ll}
    e^{\iota\varphi_{2}}|x\rangle, & \hbox{if  $\chi(x)=1$;} \\
    |x\rangle, & \hbox{if  $\chi(x)=0$.} \\
\end{array}%
\right.
\end{eqnarray}
The operators $\mathcal{P}_{0}^{\varphi_{1}}$ and
$\mathcal{P}_{\chi}^{\varphi_{2}}$ can be expressed as:
\begin{eqnarray}
\mathcal{P}_{0}^{\varphi_{1}}&=&\mathbb
I-(1-e^{\iota\varphi_{1}})|0\rangle\langle 0| \ ,\\
\mathcal{P}_{\chi}^{\varphi_{2}}&=&\mathbb
I-(1-e^{\iota\varphi_{2}})\sum_{\chi(x)=1}|x\rangle\langle x|.
\end{eqnarray}
The physical implementation of $\mathbf{Q}$ can be accomplished by
realizing the operators $\mathcal{U}$,
$\mathcal{P}_{0}^{\varphi_{1}}$ and
$\mathcal{P}_{\chi}^{\varphi_{2}}$ using external fields (or quantum
gates). The action of $\mathbf{Q}$ can be described by the following
lemma \cite{Brassard et al 1998}:

\begin{lemma}\label{Qtransformation}
Let
$\mathcal{U}|0\rangle=|\Phi\rangle=|\Phi_{g}\rangle+|\Phi_{b}\rangle$
and $g=\langle\Phi_{g}|\Phi_{g}\rangle$. Then
\begin{eqnarray}
\mathbf{Q}|\Phi_{g}\rangle&=&e^{\iota\varphi_{2}}((1-e^{\iota\varphi_{1}})g-1)|\Phi_{g}\rangle+
e^{\iota\varphi_{2}}(1-e^{\iota\varphi_{1}})g|\Phi_{b}\rangle \ ;\\
\mathbf{Q}|\Phi_{b}\rangle&=&(1-e^{\iota\varphi_{1}})(1-g)|\Phi_{g}\rangle-
((1-e^{\iota\varphi_{1}})g+e^{\iota\varphi_{1}})|\Phi_{b}\rangle.
\end{eqnarray}
\end{lemma}
From Lemma \ref{Qtransformation}, we can easily get
\begin{eqnarray}
\mathbf{Q}|\Phi\rangle&&=\mathbf{Q}(|\Phi_{g}\rangle+|\Phi_{b}\rangle)
\nonumber\\
&&=[(1-e^{\iota\varphi_{1}})(1-g+ge^{\iota\varphi_{2}})-e^{\iota\varphi_{2}}]|\Phi_{g}\rangle
+[g(1-e^{\iota\varphi_{1}})(e^{\iota\varphi_{2}}-1)-e^{\iota\varphi_{1}}]|\Phi_{b}\rangle.
\end{eqnarray}
Thus, we can amplify (or shrink) the amplitude of $|\Phi_{g}\rangle$
(or $|\Phi_{b}\rangle$) by a suitable selection of the parameters
$\varphi_{1}$, $\varphi_{2}$ in $\mathbf{Q}$ . To make this point
clearer, consider the special case $\varphi_{1}=\varphi_{2}=\pi$.
For the iteration process with
$\mathbf{Q}(\mathcal{U},\chi,\pi,\pi)$, we have the following
theorem \cite{Brassard et al 1998}:

\begin{theorem}[Amplitude Amplification]\label{AA}
Let
$\mathcal{U}|\mathbf{0}\rangle=|\Phi\rangle=|\Phi_g\rangle+|\Phi_b\rangle$,
and $\mathbf{Q}=\mathbf{Q}(\mathcal{U},\chi,\pi,\pi)$. Then, for any
$L\geq0$,
\begin{equation}
\mathbf{Q}^{L}\mathcal{U}|\mathbf{0}\rangle=\frac{1}{\sqrt{g}}\sin((2L+1)\theta)|\Phi_g\rangle
+\frac{1}{\sqrt{b}}\cos((2L+1)\theta)|\Phi_b\rangle
\end{equation}
where $b=1-g$, $\theta$ is defined so that $\sin^2\theta=g$ and
$0\leq\theta\leq\pi/2$.
\end{theorem}
Theorem \ref{AA} provides a method for boosting the initial success
probability $g=\sin^{2}\theta$. Applying the operation $\mathbf{Q}$ to the
system's initial state $L$ times, the success probability becomes
$g'=\sin^{2}((2L+1)\theta)$ and can be enhanced by choosing an integer $L$
such that $\sin^{2}((2L+1)\theta)$ is as close to $1$ as possible.

Quantum amplitude amplification has led to a quadratic speedup for
some quantum algorithms \cite{Brassard et al 1998}. In this paper we
use it as an important component for incoherent control design of
quantum systems.

\subsection{Control algorithms}
\begin{figure}
\centering
\includegraphics[width=3.6in]{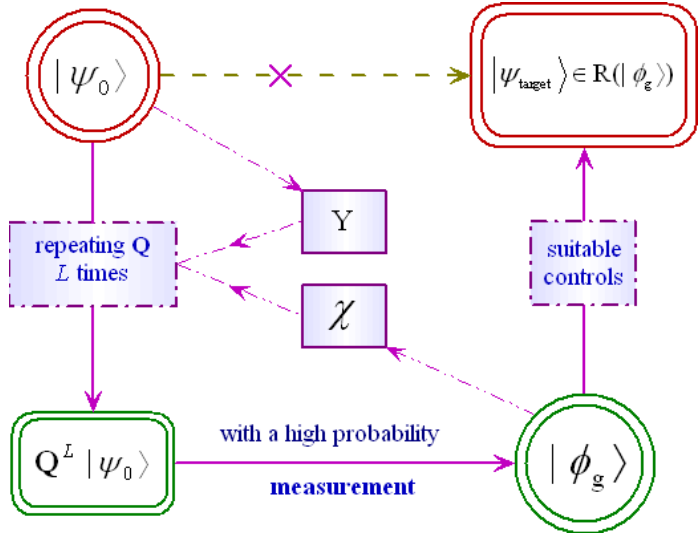}
\caption{(Color online) The schematic description of Algorithm
\ref{algorithm1}.} \label{fig algorithm1}
\end{figure}

Here we present the detailed algorithms for the proposed incoherent
control scheme. First, we assume that the target state is reachable from some eigenstate $|\phi_{g}\rangle$, i.e.,
$|\psi_{\text{target}}\rangle \in R(|\phi_{g}\rangle)$, and consider
an arbitrary initial state $|\psi_{0}\rangle$
\[
|\psi_{0}\rangle=\sum_{i=1}^{N}c_{0i}|\phi_{i}\rangle \ .
\]
We take $|\phi_{g}\rangle$ as the good state and all
$|\phi_{i}\rangle$ with $i\neq g$ as bad states, i.e.,
$|\phi_{g}\rangle$ corresponds to $\chi=1$ and $|\phi_{i}\rangle$
with $i\neq g$ corresponds to $\chi=0$. For a given initial state
$|\psi_{0}\rangle$ we can construct an operator
$\mathbf{Q}(\mathcal{U},\chi, \varphi_{1}, \varphi_{2})$ to enhance
the amplitude of getting $|\phi_{g}\rangle$, as described in
Sec.~\ref{QAA}. Here $\mathcal{U}$ is determined by the initial
state $|\psi_{0}\rangle$. Moreover, we need to find an optimal
number of iterations $L$ to maximize the probability of success.
Then we make a measurement on the enhanced conditional state such
that after the measurement the system will collapse into
$|\phi_{g}\rangle$ with a high probability. Finally we select a
suitable control to drive $|\phi_{g}\rangle$ into
$|\psi_{\text{target}}\rangle$. The main steps of the algorithm are
presented in Algorithm \ref{algorithm1} and its operation is shown
in Fig.~\ref{fig algorithm1}. The algorithm is designed for steering
the initial state $|\psi_0\rangle$ into the target state $|\psi_{\rm
target}\rangle$ which is reachable from the eigenstate
$|\phi_{g}\rangle$.

\begin{algorithm}Incoherent Control Algorithm for Eigenstate Reachable Target States\label{algorithm1}

(1) Determine a unitary operator $\mathcal{U}$ such that
$|\psi_{0}\rangle=\mathcal{U}|\phi_{1}\rangle$, and define
$|\phi_{g}\rangle$ as the good state (i.e., corresponding to
$\chi=1$) and the other eigenstates as bad states (i.e.,
corresponding to $\chi=0$);

(2) Select a pair $(\varphi_{1},\varphi_{2})$ and construct the
operation $\mathbf{Q}(\mathcal{U},\chi,\varphi_{1}, \varphi_{2})$;

(3) Find an optimal number of iterations $L$ and enhance the
amplitude of $|\phi_{g}\rangle$ by applying the operation
$\mathbf{Q}(\mathcal{U},\chi,\varphi_{1}, \varphi_{2})$ to the
system $L$ times;

(4) Make a measurement on the system, which will induce, with high
probability, collapse of the system wavefunction into the state
$|\phi_{g}\rangle$;

(5) Select a suitable control to drive $|\phi_{g}\rangle$ into
$|\psi_{\rm target}\rangle$.
\end{algorithm}

Now we consider a special class of locally controllable quantum
systems which have a wavefunction controllable subspace $\Omega$.
$\Omega$ is taken as a good subspace, i.e., the eigenstates in
$\Omega$ correspond to $\chi=1$, and we assume that the target state
$|\psi_{\text{target}}\rangle$ is reachable from the
eigenstates in $\Omega$. Without loss of generality, we assume
$D_{\omega}=\{|\phi_{1}\rangle,\dots, |\phi_{M}\rangle\}$ ($1<M<N$)
and decompose $|\psi_{0}\rangle$ into two parts:
\[
|\psi_{0}\rangle=\sum_{i=1}^{M}c_{0i}|\phi_{i}\rangle+\sum_{i=M+1}^{N}c_{0i}
|\phi_{i}\rangle=|\Phi_{g}\rangle+|\Phi_{b}\rangle \ ,
\]
where $|\Phi_{g}\rangle=\sum_{i=1}^{M}c_{0i}|\phi_{i}\rangle$ and
$|\Phi_{b}\rangle=\sum_{i=M+1}^{N}c_{0i}|\phi_{i}\rangle$. Then, an
operator $\mathbf{Q}(\mathcal{U},\chi, \varphi_{1}, \varphi_{2})$ is
constructed to enhance the amplitude of $|\Phi_{g}\rangle$. In
analogy with the Algorithm~1, below we describe the control
algorithm Algorithm~\ref{algorithm2} for steering the initial state
$|\psi_0\rangle$ of a quantum system, with a wavefunction
controllable subspace $\Omega$, into the target state $|\psi_{\rm
target}\rangle$ which is reachable from the eigenstates
in $\Omega$. A schematic of the algorithm is shown in Fig.~\ref{fig
algorithm2}.

\begin{algorithm}Incoherent Control Algorithm for Quantum
Systems with Wavefunction Controllable Subspaces\label{algorithm2}

(1) Take $\Omega$ as the good subspace such that the eigenstates in $\Omega$
correspond to $\chi=1$ and the other eigenstates correspond to $\chi=0$;

(2) Decompose the initial state $|\psi_{0}\rangle$ as the sum
$|\psi_{0}\rangle=|\Phi_g\rangle+|\Phi_b\rangle$ of the projection
$|\Phi_g\rangle$ onto the good subspace and the projection
$|\Phi_b\rangle$ onto the bad subspace;

(3) Determine $\mathcal{U}$ such that
$|\psi_{0}\rangle=\mathcal{U}|\phi_{1}\rangle$, select a pair
$(\varphi_{1},\varphi_{2})$ and construct
$\mathbf{Q}(\mathcal{U},\chi,\varphi_{1}, \varphi_{2})$;

(4) Find an optimal number of iterations $L$ and enhance the
amplitude of $|\Phi_{g}\rangle$ by applying the operation
$\mathbf{Q}(\mathcal{U},\chi,\varphi_{1}, \varphi_{2})$ to the
system $L$ times;

(5) Make a measurement on the amplified state, which will induce
collapse of the system wavefunction into a state $|\phi_{e}\rangle
\in \Omega$ with high probability;

(6) Find a local optimal control sequence in $\Omega$ and a set of
suitable controls to drive $|\phi_{e}\rangle$ into $|\psi_{{\rm
target}}\rangle$.
\end{algorithm}

\begin{figure}
\centering
\includegraphics[width=3.6in]{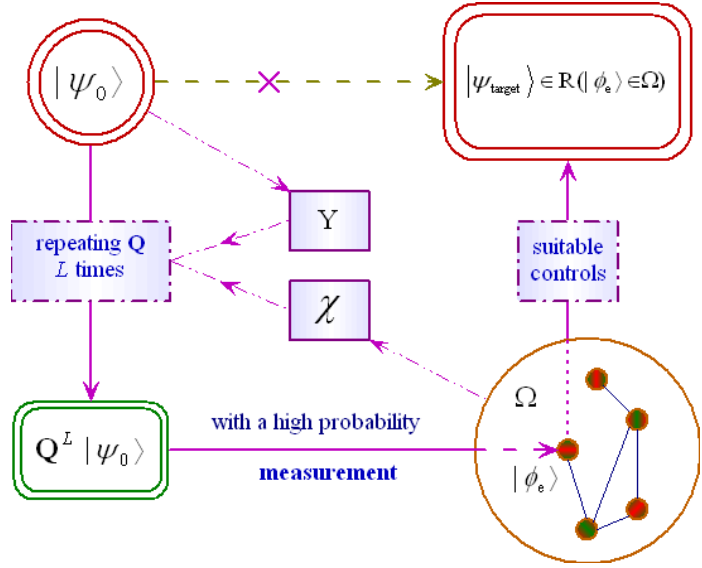}
\caption{(Color online) The schematic description of Algorithm
\ref{algorithm2}.} \label{fig algorithm2}
\end{figure}

\begin{remark}
In the proposed control algorithms,
$g=\langle\Phi_{g}|\Phi_{g}\rangle$ cannot be equal to 0. If $g=0$
for the initial state, we can make it non-zero by applying a
suitable unitary transformation to the initial state
$|\psi_{0}\rangle$ before using quantum amplitude amplification. For
globally controllable systems the amplitude amplification step is
not necessary, and the maximum probability of success is naturally
equal to 1. The selection of an optimal $\mathbf{Q}$ for the two
algorithms above remains an open issue. In Algorithm
\ref{algorithm2}, after we project the conditional state into an
eigenstate of the wavefunction controllable subspace, we can find a
local optimal control sequence by quantum reinforcement learning
\cite{Dong et al 2008}, or we can design different controllers for
every eigenstate ahead of time and complete the control task by a
variable structure control method.
\end{remark}

\begin{remark}
In the present incoherent control scheme, the measurement is used as
a control tool and the expected selectivity is enhanced at the cost
of overall conversion efficiency as the same as the case in other
measurement-assisted incoherent control schemes. For this class of
quantum systems where we know some information about local
controllability, in this paper we use quantum amplitude
amplification technology to enhance the expected selectivity and
reduce the population loss in the measurement process.
\end{remark}

\section{An example: hydrogen atom}
\begin{figure}
\centering
\includegraphics[width=3.2in]{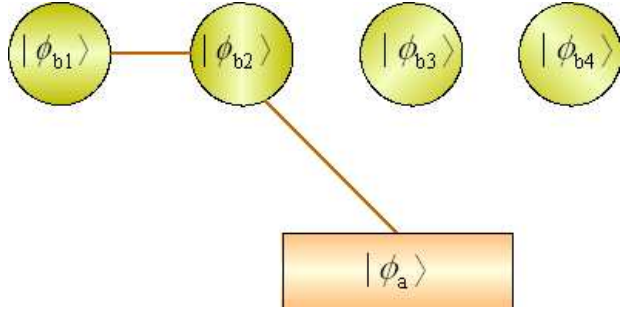}
\caption{(Color online) The state transitions of the hydrogen atom
under the action of the control Hamiltonian $H_{u}$.}
\label{hydrogen}
\end{figure}

Consider the hydrogen atom with the internal Hamiltonian
\begin{equation}
H_{0}=-\frac{\hbar^{2}}{2m}\Delta-\frac{e^{2}}{r} \ .
\end{equation}
The wavefunction of the ground state in spherical coordinates
$(r,\theta,\varphi)$ is
\begin{equation}
|\psi_{100}(r,\theta,\varphi)\rangle=\frac{1}{\sqrt{\pi
a^{3}}}e^{-r/a} \ ,
\end{equation}
where $a=0.53\times 10^{-10}m$ is the Bohr radius. The first excited state is
four-fold degenerate with the four corresponding wavefunctions
\begin{eqnarray}
|\psi_{200}(r,\theta,\varphi)\rangle&=&\frac{1}{\sqrt{8\pi
a^{3}}}(1-\frac{r}{2a})e^{-r/2a} \ ,\\
|\psi_{210}(r,\theta,\varphi)\rangle&=&\frac{1}{4\sqrt{2\pi
a^{3}}}\frac{r}{a}e^{-r/2a}\cos\theta \ ,\\
|\psi_{211}(r,\theta,\varphi)\rangle&=&-\frac{1}{8\sqrt{\pi
a^{3}}}\frac{r}{a}e^{-r/2a}\sin\theta e^{\iota\varphi} \ ,\\
|\psi_{21-1}(r,\theta,\varphi)\rangle&=&\frac{1}{8\sqrt{\pi
a^{3}}}\frac{r}{a}e^{-r/2a}\sin\theta e^{-\iota\varphi} \ .
\end{eqnarray}
We restrict consideration to these first five eigenstates and
denote them as $\{|\phi_{a}\rangle, |\phi_{b1}\rangle,
|\phi_{b2}\rangle, |\phi_{b3}\rangle, |\phi_{b4}\rangle\}$ in the
order listed above. In this model an arbitrary initial state
$|\psi_{0}\rangle$ can be written as a superposition
\begin{equation}
|\psi_{0}\rangle=c_{a}|\phi_{a}\rangle+\sum_{j=1}^{4}c_{bj}|\phi_{bj}\rangle
\end{equation}
Now consider the control design of this system. The natural simple way to control the state transitions is by using a time-dependent electric field $E_{z}(t)$ along the $z$-axis. The
corresponding control Hamiltonian $H_{u}$ has the form
$$
H_{u}=-eE_{z}(t)z \ .
$$
The wavefunction at time $t$ evolving under the action of this control Hamiltonian can be expressed as
$$|
\psi(t)\rangle=c_{a}(t)|\phi_{a}\rangle e^{-\iota E_{a}t/\hbar}
+\Bigl[\sum_{i=1}^{4}c_{bi}(t)|\phi_{bi}\rangle\Bigr]e^{-\iota
E_{b}t/\hbar} \ ,
$$
where $E_{a}$ and $E_{b}$ denote the energies of the ground and the
first excited state, respectively, and the coefficients
$C(t)=\{c_{a}(t), c_{b1}(t), c_{b2}(t), c_{b3}(t), c_{b4}(t)\}$
evolve according to the equation
$$\dot{C}(t)=TC(t)$$
Here the $T$ matrix is~\cite{Zhang 2007}:
\begin{widetext}
\begin{equation}
T=
\begin{pmatrix}
  0   & 0  & \frac{128\sqrt{2}\iota}{243\hbar}aeE_{z}(t)e^{-\iota(E_{b}-E_{a})t/\hbar}  & 0  & 0 \\
  0   & 0 & -\frac{3\iota}{\hbar}aeE_{z}(t)  & 0  & 0  \\
  \frac{128\sqrt{2}\iota}{243\hbar}aeE_{z}(t)e^{-\iota(E_{b}-E_{a})t/\hbar}   & -\frac{3\iota}{\hbar}aeE_{z}(t)  & 0 & 0  & 0 \\
  0   & 0  & 0  & 0  & 0 \\
  0   & 0  & 0  & 0  & 0 \\
\end{pmatrix} \ .
\end{equation}
\end{widetext}
It follows from the above matrix that
$$
c_{b3}(t)=c_{b3}(0),\qquad c_{b4}(t)=c_{b4}(0).
$$
Thus, the amplitudes of the states $|\phi_{b3}\rangle$ and
$|\phi_{b4}\rangle$ remain constant under the control Hamiltonian
$H_{u}$, i.e., they are non-controllable under unitary evolution.
This circumstance is similar to the example of Eqs.~(14) and~(15)
in~\cite{Turinici and Rabitz 2001}. Here the states
$|\phi_{b3}\rangle$ and $|\phi_{b4}\rangle$ are not controllable
because the field cannot effectively interact with those states due
to the selected polarization. There are no necessary coupling
interactions since the corresponding elements in
the matrix $T$ are equal to zero. The state transitions are shown in
Fig.~\ref{hydrogen}. As mentioned in Sec.~\ref{WC and WCS}, the
degeneracy condition (I) in Theorem \ref{WCS} can be relaxed in some
circumstances. Although $|\phi_{b1}\rangle$ and $|\phi_{b2}\rangle$
are degenerate, the linear span of the set
$D_{\omega}=\{|\phi_{a}\rangle, |\phi_{b1}\rangle,
|\phi_{b2}\rangle\}$ forms a wavefunction controllable subspace
$\Omega$, since the elements of $D_\omega$ are connected in
Fig.~\ref{hydrogen} \cite{Zhang 2007}. Now we consider the control
design of the hydrogen atom for two special classes of target states. The limited controllability with the Hamiltonian $H_u$ will not be enough for performing these tasks and applying the electric field along the $x$-axis or $y$-axis, or both the $x$-axis and $y$-axis will be necessary.

\textbf{Case 1: $|\psi_{\rm target}\rangle \in R(|\phi_{b4}\rangle)$}

Here we consider the situation when the target state $|\psi_{\rm target}\rangle$ is a state reachable from $|\phi_{b4}\rangle$. It can be for example an excited state above the four-degenerate first excited states (i.e., a state with the principal quantum number $n=3,4,5,\dots$). The assumption of the reachability of the target state assumes the existence of an electric control field which can transfer the state $|\phi_{b4}\rangle$ into the target state.

\begin{figure}
\centering
\includegraphics[width=4.0in]{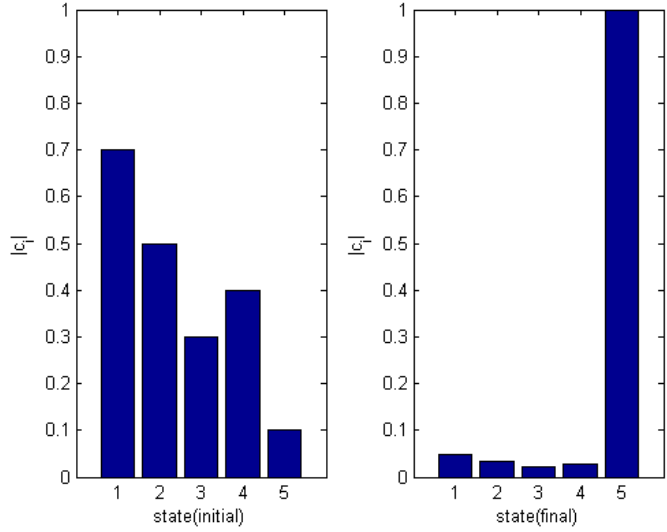}
\caption{The amplitudes of the eigenstates of the initial (left) and
final (right) states for the Case~1. The integers along the x-axis
label the hydrogen atom states.} \label{QAA1}
\end{figure}

Consider the initial state
\[
|\psi_{0}\rangle=0.7|\phi_{a}\rangle+0.5|\phi_{b1}\rangle+0.3|\phi_{b2}\rangle
+0.4|\phi_{b3}\rangle+0.1|\phi_{b4}\rangle
\]
The amplitudes of the different eigenstates are schematically shown
in the left subplot of Fig.~\ref{QAA1}. The numbers $\{1,2,3,4,5\}$
on the horizontal axis correspond to the states
$\{|\phi_{a}\rangle,|\phi_{b1}\rangle,|\phi_{b2}\rangle,
|\phi_{b3}\rangle,|\phi_{b4}\rangle\}$. The values along the
vertical axis are the corresponding
amplitudes.  If we make a measurement of the projector
$P_{b4}=|\phi_{b4}\rangle\langle\phi_{b4}|$ on the initial state,
the probability of success is only $|c_{b4}(0)|^2=1\%$, and it is
too small. Now use Algorithm \ref{algorithm1} to enhance the
probability of success. For convenience, we only use the special
quantum amplitude amplification operation
$\mathbf{Q}(\mathcal{U},\chi,\pi,\pi)$. Generation of such operator $\mathbf{Q}(\mathcal{U},\chi,\pi,\pi)$ will require applying the electric field along the $x$-axis or $y$-axis, or both the $x$-axis and $y$-axis. Using Theorem \ref{AA}, we
can compute the state of the system after repeated action of
$\mathbf{Q}(\mathcal{U},\chi,\pi,\pi)$ to the initial state seven
times. The resulting amplitudes of the final state are shown on the
right subplot of Fig.~\ref{QAA1}. Then we make a measurement of the
projector $P_{b4}$ on the final state. After the measurement the
system will collapse into the state $|\phi_{b4}\rangle$ with a high
probability of $99.53\%$. Then the last step of the incoherent control scheme is implemented by applying the coherent unitary control steering the state $|\phi_{b4}\rangle$ into the target state $|\psi_{\rm target}\rangle$.

\textbf{Case 2: $|\psi_{{\rm target}}\rangle \in R(|\phi_{e}\rangle
\in \Omega)$}

\begin{figure}
\centering
\includegraphics[width=4.0in]{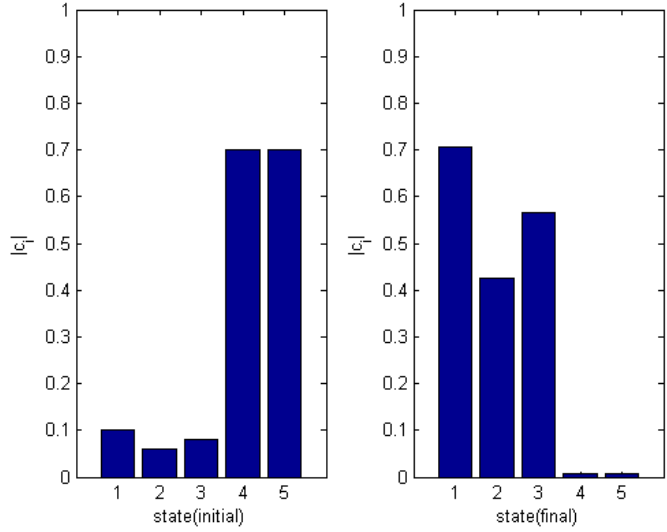}
\caption{The amplitudes of the eigenstates of the initial (left) and
final (right) states for the Case~2.}\label{QAA2}
\end{figure}

Consider the initial state
$$
|\psi_{0}\rangle=0.1|\phi_{a}\rangle+0.06|\phi_{b1}\rangle+0.08|\phi_{b2}\rangle
+0.7|\phi_{b3}\rangle+0.7|\phi_{b4}\rangle.
$$
The amplitudes of the different eigenstates are schematically shown
on the left subplot of Fig.~\ref{QAA2}. If we measure the projector
$P_\Omega$ onto the subspace $\Omega$, the probability of success
for the initial state is only
$|c_a(0)|^2+|c_{b1}(0)|^2+|c_{b2}(0)|^2=2\%$. We now use Algorithm
\ref{algorithm2} to enhance the probability of success by applying
the operation $\mathbf{Q}(\mathcal{U},\chi,\pi,\pi)$ to the system
five times. Theorem~\ref{AA} allows for computing the resulting
amplitudes, which are shown on the right subplot of Fig.~\ref{QAA2}.
Then we make a measurement on the amplified state, and the state
will collapse into subspace $\Omega$ with a high probability
$99.99\%$. After projecting the final state into the subspace
$\Omega$, we can use learning control or variable structure control
to drive the state into the target state
$|\psi_{\text{target}}\rangle$.

\begin{remark}
To demonstrate the present incoherent control scheme, we gave a
simple example of a hydrogen atom controlled by an external field.
For two special cases, we designed the corresponding amplitude
amplification operator $\mathbf{Q}(\mathcal{U},\chi,\pi,\pi)$ (a
unitary transformation). The unitary transformation can be
constructed using the method presented in Section III.B according to
nature of the initial states and local controllability information.
One goal of this work is to establish a connection between control
design and controllability analysis, and also demonstrate that local
controllability information can be used to design control laws for
quantum systems. Further, it is necessary to generate effective
external control field to realize the known unitary transformation
$\mathbf{Q}(\mathcal{U},\chi,\pi,\pi)$. In this example, the control
field should include some components along some other axes (e.g.,
$x$-axis, $y$-axis, or both the $x$-axis and $y$-axis). The details
of control field design are left to future work. Moreover, our
control scheme can easily be applied to multiple level systems such
as the five level system in \cite{Tersigni et al 1990}. If we know
some information about local controllability and there are fewer
nonzero component values in the electric dipole transition matrix
than those given in \cite{Tersigni et al 1990} and \cite{Turinici
and Rabitz 2001}, we can use a method similar to the one used in
this example to design the control law.
\end{remark}

\section{Concluding remarks}
Many of the works studying the controllability of quantum systems are devoted
to the analysis of global properties of controlled quantum systems. However,
some quantum systems may not be globally controllable or sometimes information
about global controllability may be difficult to deduce, and we can then consider operation with locally controllable quantum systems. In fact, a system with a
controllable subspace is a special case of locally controllable systems.
Systems with controllable subspaces have practical importance in quantum
information technology~\cite{Kiffner et al 2007}, e.g., systems with a DFS in quantum computation~\cite{Lidar et al
1998}. It is valuable to steer the system into a DFS and to maintain the
encoded information in the DFS for the computation operations through
suitable control schemes \cite{Kiffner et al 2007,Cappellaro et al 2006}. We
believe that the present incoherent control scheme can provide an alternative
method for such problems.

In the present control scheme, quantum measurement is used as an
effective control to probabilistically project the amplified state
into a desired eigenstate. This scheme uses controlled unitary
transformations supplemented by quantum measurements and thus can be
considered within the class of measurement-assisted
schemes~\cite{Pechen et al 2006,Shuang et al 2007}. Since the
quantum measurement makes the state collapse probabilistically, our
incoherent control scheme is also a probabilistic control strategy,
similar in some sense to the incoherent control scheme of~\cite{Dong
et al 2007}. However, the methods of enhancing the probability of
success are quite different. In \cite{Dong et al 2007}, the
probability of success is enhanced by multiple measurements on
identical initial states. Generally it requires several identical
initial states but they may be unknown. In the present scheme,
quantum amplitude amplification is used to increase the probability
of success, and the initial state should be known. In this sense,
the present scheme is similar to the quantum control scheme based on
the use of Grover iterations~\cite{Zhang et al 2005}. However, the
quantum control algorithm in~\cite{Zhang et al 2005} relies on the
assumption that the system is eigenstate controllable, which may be
difficult to verify. Moreover, the construction of Grover iterations
also has some special requirements, which may not be suitable for
control design of general quantum systems. The present scheme
overcomes these drawbacks and is more suitable for accomplishing
practical control tasks. Here we only require the system to be
locally controllable and use the general amplitude amplification
technique to increase the probability of success. The amplitude
amplification operations can be directly realized by the
construction of a suitable control Hamiltonian. It is obvious that
this incoherent control scheme is also suitable for the control of
globally controllable quantum systems.

In conclusion, we propose an incoherent control scheme for a class
of quantum systems exhibiting local controllability. In this scheme,
we only need to know partial controllability information about
the system. The paper also provides an example with the hydrogen atom
to demonstrate the operation of the scheme. In our method, quantum
measurements are used as an effective control and the gap between
a controllability analysis and control design is bridged. The
experimental details for implementing the scheme need to be
explored for the practical manipulation of quantum systems.

\begin{acknowledgments}
The authors would like to thank the reviewer of J. Chem. Phys. for
helpful comments which have greatly improved this paper. D.D. would
like to thank Prof. Lei Guo and Dr. Bo Qi for helpful suggestions.
This work was supported in part by the National Natural Science
Foundation of China under Grant No.60703083, the National Creative
Research Groups Science Foundation of China under Grant No.60421002
and the China Postdoctoral Science Foundation (20060400515). H.
Rabitz and A. Pechen thank DOE for support. A. Pechen also thanks
RFFI 08-01-00727-a for a partial support.
\end{acknowledgments}

\newpage 
\bibliography{apssamp}

\end{document}